\documentclass[useAMS,usenatbib]{mn2e}

\usepackage{psfig} 
\usepackage[dvips]{graphicx}

\def\rxte{{\sl RXTE}}
\def\mch{M$\rm^{c}$Hardy}

\def\ngc4051{NGC~4051}
\def\mkn79{Mrk~79}
\def\msun{$M_\odot\,$}
\newcommand{\be}{\begin{equation}}
\newcommand{\ee}{\end{equation}}

\title[X-ray/optical variability in NGC~4051]{Twelve years of X-ray and optical variability in the Seyfert galaxy NGC~4051}
\author[E. Breedt et al.]{E. Breedt$^{1}$\thanks{E-mail: ebreedt@astro.soton.ac.uk}, I. M. \mch$^{1}$, P. Ar\'evalo$^{2,3}$, P. Uttley$^{1}$, S. G. Sergeev$^{4,5}$, \and T. Minezaki$^{6}$, Y. Yoshii$^{6,7}$, Y. Sakata$^{8}$, P. Lira$^{9}$, N. G. Chesnok$^{10,11}$\\  
$^1$School of Physics and Astronomy, University of Southampton, Southampton, SO17 1BJ, UK\\
$^2$Shanghai Astronomical Observatory, 80 Nandan Road, Shanghai 200030, China\\
$^3$Departamento de Ciencias Fisicas, Universidad Andres Bello, Av. Republica 252, Santiago, Chile\\
$^4$Crimean Astrophysical Observatory, P/O Nauchny, Crimea 98409, Ukraine\\
$^5$Isaac Newton Institute of Chile, Crimean Branch, Ukraine\\
$^6$Institute of Astronomy, School of Science, University of Tokyo, 2-21-1
Osawa, Mitaka, Tokyo 181-0015, Japan\\
$^7$Research Centre for the Early Universe, School of Science, University of
Tokyo, 7-3-1 Hongo, Bunkyo-ku, Tokyo 113-0033, Japan\\
$^8$Department of Astronomy, School of Science, University of
Tokyo, 7-3-1 Hongo, Bunkyo-ku, Tokyo 113-0033, Japan\\
$^{9}$Departamento de Astronom\'{i}a, Universidad de Chile, Casilla 36-D, Santiago, Chile\\
$^{10}$Taras Shevchenko National University of Kiev,
pr. Akademika Glushkova 2, Kiev, 03680 Ukraine\\
$^{11}$Main Astronomical Observatory, Ukrainian National Academy of Sciences, ul. Akademika Zabolotnogo, 27, Kiev, 03680 Ukraine
}

\begin{document}

\date{Accepted/Received}

\pagerange{\pageref{firstpage}--\pageref{lastpage}} \pubyear{2009}

\maketitle

\label{firstpage}

\begin{abstract}
We discuss the origin of the optical variations in the Narrow line Seyfert~1 galaxy NGC~4051 and present the results of a cross-correlation study using X-ray and optical light curves spanning more than 12~years. The emission is highly variable in all wavebands, and the amplitude of the optical variations is found to be smaller than that of the X-rays, even after correcting for the contaminating host galaxy flux falling inside the photometric aperture. The optical power spectrum is best described by an unbroken power law model with slope $\alpha=1.4^{+0.6}_{-0.2}$ and displays lower variability power than the 2--10~keV X-rays on all time-scales probed. We find the light curves to be significantly correlated at an optical delay of $1.2^{+1.0}_{-0.3}$~days behind the X-rays. This time-scale is consistent with the light travel time to the optical emitting region of the accretion disc, suggesting that the optical variations are driven by X-ray reprocessing. We show, however, that a model whereby the optical variations arise from reprocessing by a flat accretion disc cannot account for all the optical variability. There is also a second significant peak in the cross-correlation function, at an optical delay of $39^{+2.7}_{-8.4}$~days. The lag is consistent with the dust sublimation radius in this source, suggesting that there is a  measurable amount of optical flux coming from the dust torus. We discuss the origin of the additional optical flux in terms of reprocessing of X-rays and reflection of optical light by the dust.

\end{abstract}


\section{Introduction} \label{sec:intro}

The exact mechanism producing the \mbox{X-ray} emission from active galactic nuclei (AGN) is not well understood, although most models involve Comptonization of the UV/optical spectrum, through a tenuous, hot corona above the inner disc  \citep{haardtmaraschi91}. The fastest variability time-scales set a limit on the size of the emission region, suggesting that the \mbox{X-rays} are emitted from the innermost regions of the system, but the exact geometry is not known.
The optical continuum is variable as well, and thought to be thermal emission from an optically thick accretion disc, surrounding the black hole \citep{koratkarblaes99}. The origin of the variability, as well as the connection between the disc and the corona, is still unclear, however. It has been suggested that the short time-scale optical variations are the result of reprocessing of the highly variable \mbox{X-rays} by the disc \citep{krolik91}, enhancing the intrinsic thermal emission from the disc. Hence variations in the \mbox{X-rays} should be mirrored in the optical light curves, smoothed by the large size of the disc and delayed by the light travel time between the source and the disc. %
Although there are some sources that do show evidence of reprocessing \citep[e.g.][]{wandersetal97, collier98, oknyanskij03, sergeevetal05, cackettetal07}, the results are not generally applicable. Many sources show no simple relationship between bands \citep{nandra00,gaskell08} and at least one source shows no correlation at all \citep{maoz00,maoz02}.
An increasing number of sources display optical variations of greater amplitude than their corresponding \mbox{X-ray} variations, such as NGC~5548 \citep{uttley03}, MR2251-178 \citep{arevalo08}, \mkn79 \citep{breedt09mkn79} and NGC~3783 \citep{arevalo09}, making it difficult to attribute all the optical variability to reprocessing. Energetics arguments also rule out reprocessing as the main driver of the variability in sources where the variable optical luminosity exceeds the \mbox{X-ray} luminosity \citep{gaskell07}. 

Also important to consider, is variability intrinsic to the disc, such as instabilities propagating radially inward through the disc, affecting the mass accretion rate and local emission \citep[e.g.][]{lyubarskii97}. In this case, optical variations are expected to precede similar variations in the \mbox{X-rays}. The time-scales involved are much longer, as variations will propagate on the viscous time-scale, which is $\sim$months--years at the optical emitting region in AGN.
The long term ($\sim$months--years) optical variability has been explained by composite models including accretion rate fluctuations propagating through the disc \citep{arevalouttley06,arevalo08} and changes in the geometry of the system \citep{breedt09mkn79}. Reprocessed emission from the broad line region may also contribute to the optical variations \citep{arevalo09,koristagoad01}, and \citet{gaskell07} suggested that there might be a contribution to the optical emission from reprocessing material much further out, such as the molecular torus. 

By undertaking simultaneous multi-wavelength monitoring and studying the correlation between the emission at different wavelengths, we can determine the dominant process underlying the optical variations and gain some physical insight into the connection between the emission regions.

\citet{uttley03} proposed that the dominant process driving the observed optical variability, was dependent on the mass and accretion rate of the black hole, as these parameters define the temperature profile of the accretion disc. Within the standard accretion disc model \citep[e.g.][]{shakura}, AGN with smaller black hole mass and/or higher accretion rate, will have have hotter discs than systems of high mass/low accretion rate. Scaled in terms of the gravitational radius, the \mbox{X-ray} and optical emitting regions are further apart in the low mass systems, so the viscous time-scale in the optical emitting region is very long compared to that in the inner disc. We may therefore expect reprocessing to be the main contributor to the fast optical variations, and accretion rate variations to occur on time-scales of several tens of years.

NGC~4051 is a low-luminosity narrow-line Seyfert~1 galaxy, known for its extreme \mbox{X-ray} variability \citep[e.g.][]{mchardy04psd4051}. 
The relationship between the \mbox{X-ray} and UV/optical variability in this galaxy has been the subject of many papers, reporting varied and conflicting results.  
Over the course of 3 nights \citet{done90} found the \mbox{X-rays} to vary by a factor of 2, but no corresponding variations in the optical or infrared were seen. 
\citet{klimek04} also studied the short time-scale optical variability of this galaxy. They detected microvariability on one of the five nights it was observed. 
\citet{uttley00} found a strong correlation between the \mbox{X-ray} and EUV variations, and measured them to be simultaneous to within 1 ks. \citet{peterson00} concluded that the \mbox{X-ray} and optical emission was correlated on time scales of months to weeks, but by applying a 30 day smoothing boxcar to the light curves before calculating the cross-correlation, a measurement of any lag shorter than this could not be obtained. During a period of very low \mbox{X-ray} flux they found that the broad He {\sc ii} $\lambda4686$ line greatly decreased in intensity as well, showing no noticeable variation during this period, but that the optical continuum and H$\beta$ line were only slightly fainter and continued to vary. They interpreted this is as the inner disc turning into an advection-dominated flow, greatly reducing the short wavelength (UV/blue) emission from this part of the disc.
A 3-month simultaneous \mbox{X-ray} and optical monitoring program by \citet{shemmer03} revealed significant evidence of a correlation close to zero lag. They  found the optical emission in this case to {\em lead} the \mbox{X-ray} emission by 2.4 days, and note that the variability of the optical light curve was much less than that of the \mbox{X-rays}. On shorter time-scales, \citet{mason02} find the UV to lag the \mbox{X-rays}, which they interpret as reprocessing by a ring of material at a distance of 0.14~light-days. This result has also been confirmed by \citet{smithvaughan06}, using the same data.
Reverberation mapping studies using the H$\beta$ line \citep{peterson00, kaspi00, peterson04} place the broad line region in this AGN at a distance of $\sim$6 light-days from the central ionising source.

It is clear that a combination of processes can contribute to the observed optical variability and that different processes can drive the variability on different time-scales. In this paper, we combine the observations from 4 different telescopes with some archival data, to revisit the topic of \mbox{X-ray} and optical variability in NGC~4051. The light curves together span more than 12~years, allowing us to study the variability on much longer time-scales than has been done before in this source. 
This paper is the fourth in a series describing our findings from a sample of Seyfert galaxies monitored at \mbox{X-ray} and optical wavelengths. Previously we have reported on our results on MR2251-178 \citep{arevalo08}, \mkn79 \citep{breedt09mkn79} and NGC~3783 \citep{arevalo09} and here we discuss our findings on NGC~4051, the lowest mass ($M=(1.91\pm0.78)\times10^6$\msun, \citealt{peterson04}), lowest luminosity ($L_{\mbox{\tiny{bol}}}<10^{44}$ erg s$^{-1}$) galaxy in our sample. 
Throughout we adopt the average Tully-Fisher distance estimate of 15.2~Mpc \citep{russell03} to this source.

In Section~\ref{sec:data} we present the multicolour optical and \mbox{X-ray} data. We show the optical power spectrum in Section~\ref{sec:powerspectra} and in Section~\ref{sec:ccf} we present the results of a cross-correlation analysis of the light curves. We discuss the results in terms of a simple reprocessing model in Section~\ref{sec:repr} and show that there is an additional component to the optical variability. We show that these additional variations are delayed with respect to the \mbox{X-rays} by a time-scale consistent with the light travel time to the torus, so in Section~\ref{sec:discussion} we investigate the possibility of optical emission originating from the torus, in detail. Our findings are summarised in Section~\ref{sec:conclusion}.
%-----------------------------------------------------------------------------------------
\section{Data} \label{sec:data}
%-----------------------------------------------------------------------------------------

%--------------------------------------V-BAND-LIGHTCURVE----------------------------------
\subsection{V band optical observations}
\label{sec:Vlc}
The $V$ band optical light curve presented in this paper is a combined light curve from four telescopes: the Liverpool Telescope (LT) at the Observatorio del Roque de Los Muchachos of the Instituto de Astrofisica de Canarias, La Palma, Spain\footnote{The Liverpool Telescope is operated by Liverpool John Moores University with financial support from the UK Science and Technology Facilities Council.} \citep{livtel}, the Faulkes Telescope North (FT) at Haleakala, Maui, Hawaii, the MAGNUM (Multicolor Active Galactic Nuclei Monitoring) telescope \citep{yoshii03magnum} at the University of Hawaii's Haleakala Observatory, Maui, Hawaii, and the 0.7-m telescope of the Crimean Astrophysical Observatory (CrAO) in the Crimea, Ukraine \citep{sergeevetal05,chesnok09}.
The data reduction procedures and light curve construction are described in detail in \citet{breedt09mkn79}, so we only give a brief summary below.

We observed NGC~4051 through a Bessell-$V$ band filter from 2004 December~9 to 2008 December~11 with the Liverpool Telescope and from 2007 February~5 to 2007 August~10 with the Faulkes Telescope North. The two telescopes are identical in design. Observations in other filter bands, on both telescopes, started in early 2007 and continued until 2007 August on the FT and 2008 July on the LT.  The data reduction and photometry was done using standard techniques in {\sc iraf}. We performed aperture photometry on the individual images, using an aperture of 12\arcsec$\,$ diameter, centred on the nucleus of the galaxy. The relative fluxes were calibrated with respect to four comparison stars in the field, and the final flux calculated using the comparison star magnitudes of \citet{doroshenko05i}.  

The MAGNUM telescope observed NGC~4051 in the $V$ band from 2001 January 8 to 2007 July 30. Data from the start of this program until MJD 52819 (2003 June 29) have been published by \citet{suganumaetal06}. The rest of the light curve was previously unpublished.  

Observations at the CrAO were carried out between 2001 December~27 and 2006 August~18. The filter set on the Crimean telescope is non-standard, but their $B,V,R$ and $I$ filters are close to the Johnson/Bessell filters and their $R1$ filter is similar to the Cousins $I$ filter \citep[see][]{doroshenko05i}.

The light curves from the different telescopes were made using different aperture sizes, so each is affected by a different amount of host galaxy contamination, resulting in different average flux levels.  To account for this, as well as for the slight differences in the filters and any other small systematic differences in the calibration, we use the overlapping parts of the light curves to match them in a least squares fashion. We extracted pairs of measurements, taken within one day of each other, from overlapping light curves, and performed a least-squares fit of the equation $F_1 = aF_2 + b$ to the flux pairs. $F_1$ and $F_2$ are the flux measurements from the two datasets and $a$ and $b$ the parameters of the fit. The points were weighted for the fit, by the inverse of the time difference between $F_1$ and $F_2$, so that measurements made closer together in time have a greater influence on the fit. The best-fit parameters from each pair of overlapping light curves were then used to scale that dataset to the 15\arcsec$\,$ diameter aperture $V$~band light curve from the CrAO telescope. 

\begin{figure*}
\includegraphics[width=16.5cm]{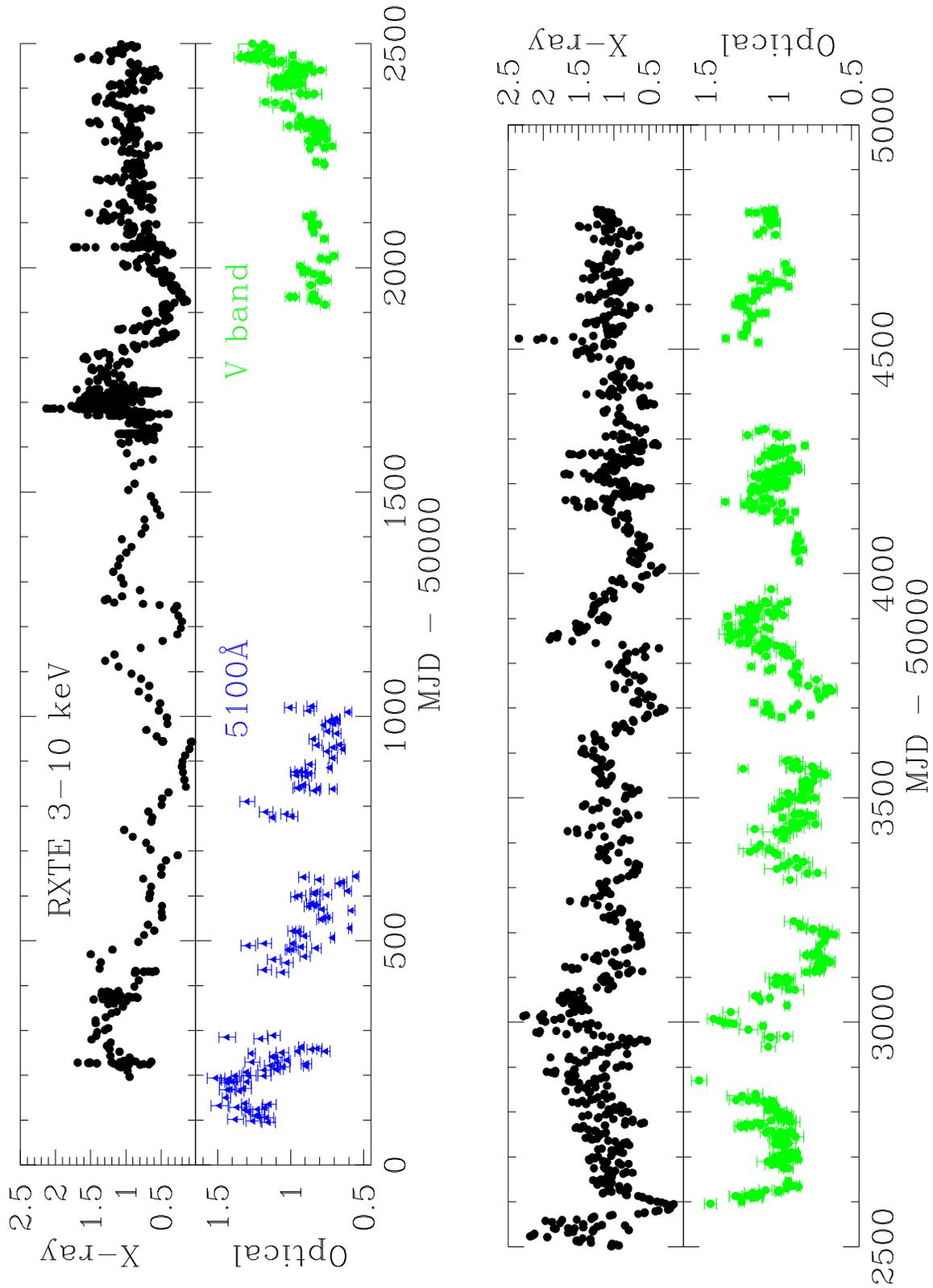}
\caption{$3-10$ keV X-ray (top panel in each part) and optical (bottom panel in each part) light curves of NGC~4051, corrected for host galaxy flux inside the aperture. Each light curve is normalised to its mean flux for direct comparison of the variability amplitudes. Note that the vertical scales for the optical and X-ray light curves are different -- the amplitude of the optical variations is smaller than that of the X-rays. The X-ray light curve is smoothed with a 4 point ($\sim$1~week) running average (for plotting purposes only), to highlight the variations on time-scales similar to that of the optical light curve. The first part of the optical light curve, plotted in blue triangles, is the 5100\AA$\,$ continuum light curve from the AGN~Watch program. The rest of the light curve, plotted in green dots, is the combined $V$ band light curve from 4 different telescopes.\label{fig:xvlc}}
\end{figure*}

The $V$~band light curve used in the analysis, is the final combined light curve, corrected for Galactic reddening and host galaxy flux inside the aperture (see Section~\ref{sec:galaxyflux}). It is shown in Figure~\ref{fig:xvlc}, normalised to its mean flux, for direct comparison with the \mbox{X-ray} variations. Full sampling characteristics of this light curve may be found in Table~\ref{tab:lcprop}.

%-----------------------------------------AGN-WATCH----------------------------------------
\subsection{AGN Watch continuum light curve}
NGC~4051 was also observed spectroscopically between 1996 January 12 and 1998 July 28 as part of the International AGN Watch program \citep[e.g.][]{peterson99agnw}. 
Light curves from this program are publicly available\footnote{http://www.astronomy.ohio-state.edu/$\sim$agnwatch/} and we include in our analysis the 5100\AA$\,$ continuum light curve originally presented by \citet{peterson00}. We subtracted the host galaxy contribution inside the 5.0\arcsec$\times$7.5\arcsec$\,$ spectroscopic aperture, as measured by \citet{bentz08}, and converted the light curve to mJy units, to compare with our $V$ band light curve. Note that in this light curve the continuum flux is measured from a line-free region around 5100\AA$\,$ of the optical spectra, while the broad band observations will contain the line emission that fall inside the $V$ band bandpass, such as [O{\sc iii}]$\lambda\lambda$4959,5007 and H$\beta\,\lambda$4861. We discuss the contribution of the lines to the flux in Section~\ref{sec:n4051ccfxrayopt}.
The light curve is plotted in Figure~\ref{fig:xvlc}, together with the X-ray and $V$ band light curves.

%--------------------------------------COLOUR-LIGHTCURVES----------------------------------
\subsection{Multicolour optical observations}
\label{sec:colourlcs}
\begin{figure*}
\rotatebox{270}{\includegraphics[width=14cm]{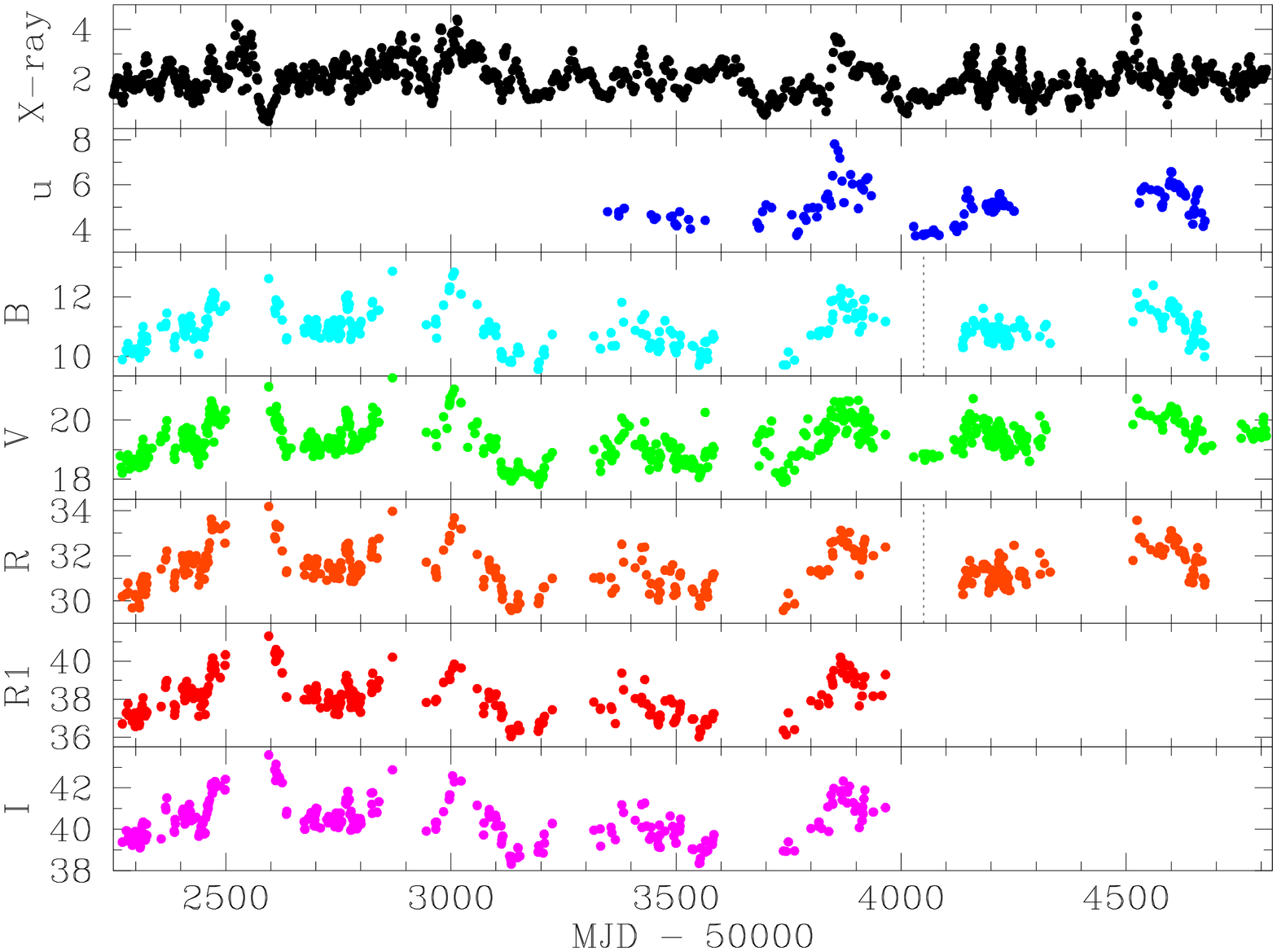}}
\caption{X-ray, $u$, $B$, $V$, $R$, $R1$ and $I$ band light curves of NGC~4051. The X-ray light curve is in units of $10^{-11}$~erg~s$^{-1}$~cm$^{-2}$ and the optical light curves are in mJy. The fluxes have been corrected for the small amount of Galactic extinction towards the source, but host galaxy flux has not been subtracted. The dotted lines in the $B$ and $R$ band light curves indicate the positions where there might be a small offset between the light curves on either side of the line, as there was no overlap between the data from the different telescopes. This does not affect our cross-correlation results however (see text).}
\label{fig:colourlcs}
\end{figure*}

%-----------------

\begin{table*}
   \caption{Observational characteristics of all light curves \label{tab:lcprop}}
   \begin{tabular}{c c c c c c c c}
      \hline\hline
      Light curve & Telescope & Observation dates & Total length & Number  & $\Delta t_{\mbox{\tiny mean}}$ & $\Delta t_{\mbox{\tiny median}}$ & Mean flux$^{*}$ \\
      ~ & ~ & (MJD) & (d) & of epochs & (d) & (d) & (mJy) \\
      \hline
      X-ray     & \em{RXTE}     & $50196-54811$ & 4615 & 1811 & 2.55 & 1.96 & $1.93\pm0.95^{\ddagger}$\\
      $u$       & LT            & $53348-54675$ & 1327 &  128 & 10.4 & 4.01 & $5.41\pm0.93$\\
      $B$       & combined      & $52271-54675$ & 2404 &  386 & 6.24 & 1.96 & $11.5\pm0.61$\\
      $5100$\AA & AGN$\,$Watch  & $50095-51023$ &  928 &  126 & 7.42 & 3.15 & $11.1\pm0.93$\\
      $V$       & combined      & $51917-54811$ & 2894 &  629 & 4.61 & 1.94 & $20.1\pm0.64$\\
      $R$       & combined      & $52271-54675$ & 2404 &  384 & 6.28 & 1.97 & $32.3\pm0.91$\\
      $R1^{\dagger}$& CrAO      & $52271-53965$ & 1694 &  265 & 6.42 & 1.98 & $38.9\pm1.02$\\
      $I$       & CrAO          & $52271-53965$ & 1694 &  261 & 6.52 & 1.98 & $41.3\pm1.02$\\       	
      \hline
   \end{tabular}
   \begin{flushleft}   
      \small{$^{*}$ Including host galaxy flux}\\
      \small{$^{\ddagger}$ In units of $10^{-11}$ erg s$^{-1}$ cm$^{-2}$}\\
      \small{$^{\dagger}$ The $R1$ filter corresponds approximately to the Cousins $I$ filter}
   \end{flushleft}   
\end{table*}

%-----------------

In addition to the $V$ band observations described above, the CrAO also observed NGC~4051 in the $B$, $R$, $R1$ and $I$ filter bands during the same period. The light curves are constructed from aperture photometry through a $15\arcsec$ diameter aperture centred on the galaxy nucleus, relative to a comparison star in the field. We use the comparison star fluxes from \citet{cackettetal07} to convert the light curves to flux units in these bands. The data from the start of this program, until MJD 53022 (2004 January 18), have been published by \citet{sergeevetal05} and were also analysed by \citet{cackettetal07}. Data after this date were previously unpublished.

We supplement these light curves with Sloan Digital Sky Survey (SDSS)-$u$ band observations made by the Liverpool Telescope between 2004 December 9 and 2008 July 27. The data from short monitoring programs in $B$ and $R$ on the Liverpool and Faulkes telescopes (2007 February 5 -- 2008 July 27)  were combined to extend the CrAO light curves. 
Unfortunately there is no overlap between the LT/FT and the CrAO light curves in these bands, so it was not possible to combine them as described in section~\ref{sec:Vlc}. To ensure the best possible match between the light curves, the same comparison stars were used in the flux calibration of the two datasets, and the appropriate aperture corrections were applied to the LT/FT light curves. However, it is possible that small differences in the calibration, or between the filters on the different telescopes, may cause a small offset between the CrAO and LT/FT light curves. The positions in the light curves that are affected by this, are indicated by dotted lines in Figure~\ref{fig:colourlcs}. Note however, that due to the ``segmentation'' technique we use to calculate the cross-correlation (Section~\ref{sec:ccf}), a possible small offset like this does not affect our results.
The light curves are shown, together with the corresponding segment of the X-ray light curve, in Figure~\ref{fig:colourlcs}. 

The flux uncertainty on each point, based upon the differential photometry error and the relative calibration of the light curves, is approximately $3\%$. The uncertainty in the comparison star fluxes introduce a $1-4\%$ error in average flux of the different bands, but note that this error does not increase the scatter in the light curve, so it does not affect the cross-correlation results below. Sampling statistics for each light curve are shown in Table~\ref{tab:lcprop}.

%--------------------------------------X_RAY-OBSERVATIONS-----------------------------------
\subsection{X-ray observations}
NGC~4051 is part of our long term AGN \mbox{X-ray} monitoring program using \rxte, which started on 1996 April 23. Here we include the observations up to 2008 December 11, to overlap with the optical light curve. The observational frequency and variability characteristics of this light curve are summarised in Tables~\ref{tab:lcprop} and~~\ref{tab:galflux}.
The observations are in the form of approximately 1~ks snapshots using the Proportional Counter Array (PCA) on board \rxte. The PCA consists of five identical proportional counter 
units (PCUs) but as some of them were regularly switched off during our observations, we only extracted data from PCU2 to construct the \mbox{X-ray} light curve. For maximum signal-to-noise, we used only the data from the top layer. Using {\sc ftools} v.6.4, we applied standard extraction techniques and re-reduced all the available data, so that the latest \rxte$\;$ background models may be applied throughout the light curve. We rejected data obtained within 10~minutes of passing through the South Atlantic Anomaly (SAA), data gathered less than 10\degr$\,$ above the limb of the Earth or with pointing offset greater than 0.02\degr$\,$ from the source, and data for which the electron current was greater than 10\%. To obtain the flux measurement, we fit a power law model to the spectra using {\sc xspec} and integrate the flux in the range 3--10~keV. The error on the flux measurements were calculated from the counts in each observed spectrum, weighted by the response function.

We show the normalised 3--10~keV \mbox{X-ray} light curve in Figure~\ref{fig:xvlc}, smoothed with a 4-point ($\sim$1 week) running average.  This is done to highlight fluctuations on similar time-scales as those in the optical light curve, but note that, throughout this paper, the unsmoothed, unbinned light curve was used in the analysis.

%--------------------------------------HOST-GALAXY-FLUX----------------------------------
\subsection{Galaxy flux in the aperture} \label{sec:galaxyflux}
The photometry aperture contains the nuclear emission as well as a constant contribution from stars in the host galaxy. Although this constant offset does not affect the results of the cross-correlation analysis, it reduces the amplitude of the observed variations.  Hence, to investigate the variability of the nucleus, this component must be subtracted from the light curve.     

In order to estimate the amount of host galaxy flux contained in the 15\arcsec$\,$ aperture, we decomposed the images using {\sc galfit} \citep{peng02_galfit}, by fitting a psf (point spread function -- using stars in the field of the galaxy), a de Vaucouleur profile and an exponential disc to each image. The decomposition of individual images was poor, so we stacked together 12 images taken under good seeing conditions, to improve the signal-to-noise. This resulted in much better, stable fits of the different components. The variable nucleus was taken account of by selecting only images from which we measured approximately the same flux through the aperture. We subtracted the best fit galaxy profile in each band from the images and measured the flux of the residual psf-like image. The difference between this psf-like image and the flux measured through a 15\arcsec$\,$ aperture on the original image, was taken as the galaxy contribution $f_g$. We list these fluxes, corrected for foreground Galactic extinction, in Table~\ref{tab:galflux}. 

\begin{figure}
\rotatebox{270}{\includegraphics[width=6.5cm]{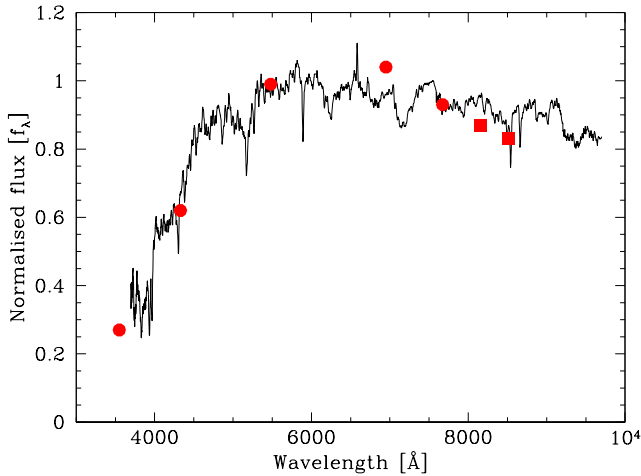}}
\caption{Measured galaxy flux inside the aperture (plotted in dots), compared to a template spectrum of the central regions of a normal spiral galaxy similar to NGC~4051. The last two points, plotted in squares, are the flux estimates for the $R1$ and $I$ bands. These points were extrapolated from the SDSS-$i$ flux measurement (last dot), as we did not have images available in these bands.}
\label{fig:galspectrum}
\end{figure}

We did not have $R1$ and $I$ band images available to measure the galaxy contribution by decomposition as above, so the galaxy contribution in these bands were estimated by extrapolating from the galaxy contribution measured in the SDSS-$i$ band. 
The fluxes in Table~\ref{tab:galflux} are in good agreement with those found by \citet{salvati93} and also agree well with the galaxy-subtracted $V$ band light curve published by \citet{suganumaetal06}.
We show in Figure~\ref{fig:galspectrum} the measured galaxy fluxes, converted to $F_\lambda$ units and corrected for the small amount of foreground Galactic extinction towards the source. It is shown together with a template spectrum of the central regions of a normal spiral galaxy similar to NGC~4051 \citep{santos02_spectra,bica88_galspec}. The template spectrum was constructed from a  group of 30 galaxies, mostly of morphological type Sb, but also including Sa and Sc galaxies. NGC~4051 is classified as an Sbc galaxy. The template is corrected for Galactic extinction and normalised at 5870\AA. Our galaxy fluxes agree well with the template spectrum, indicating that our $f_g$ measurements are good estimates of the host galaxy flux inside the aperture. Flux-flux plots between galaxy-subtracted light curves in different energy bands suggest that the error on $f_g$ is between 10 and 15\%.

%------------------------------------FRACTIONAL VARIABILITY------------------------------
\subsection{Fractional variability}

The fractional variability $F_{\mbox{\tiny var}}$ describes the flux variance of the light curve as a fraction of the mean flux $\langle f \rangle$, and provides a way of comparing the amount of intrinsic variability in each wavelength band. Mathematically,
\be F_{\mbox{\tiny var}}= \frac{\sqrt{\sigma^2-\langle \varepsilon^2 \rangle}}{\langle f \rangle}\;. \label{eq:Fvar} \ee%
Here $\sigma$ is the flux variance %
\be \sigma^2 = \frac{1}{N}\sum_{i=1}^{N}(f_i-\langle f \rangle)^2 \ee %
and $\langle \varepsilon^2 \rangle$ is the mean of the squared measurement errors
\be \langle \varepsilon^2 \rangle = \frac{1}{N}\sum_{i=1}^{N}\varepsilon_i^2. \label{eq:errsqr}\ee %
The error on $F_{\mbox{\tiny var}}$ is given by
\be \varepsilon_{(F_{\mbox{\tiny var}})} = \sqrt{ \left(\sqrt{\frac{1}{2N}}\;\frac{\langle \varepsilon^2 \rangle}{\langle f \rangle^2 F_{\mbox{\tiny var}}}\right)^2  +  \left( \sqrt{\frac{\langle \varepsilon^2 \rangle}{N}}\;\frac{1}{\langle f \rangle}\right)^2}  \label{eq:errFvar} \ee %
\citep{vaughan03}. This error describes only the error on $F_{\mbox{\tiny var}}$ due to observational noise. In our case, there is also an uncertainty associated with the galaxy subtraction. This uncertainty does not affect the scatter of the points in the light curve (since we are subtracting the same galaxy contribution to each flux measurement), so should not be included in the average error in Eqn.~\ref{eq:errsqr}. It should be accounted for in the error on $F_{\mbox{\tiny var}}$ however, so we also include a third error term in the square root of Eqn.~\ref{eq:errFvar} of the form $(F_{\mbox{\tiny var}}\:\Delta\langle f \rangle/\langle f \rangle\:)^2$, where $\Delta\langle f \rangle$ is the error in the galaxy flux contribution.

The fractional variability, as well as the ratio of the maximum to minimum flux in each light curve, is given in Table~\ref{tab:galflux}. We also list the variability characteristics of the 5100\AA$\,$ continuum light curve together with the galaxy contribution measured by \citet{bentz08}, but point out that this light curve was made using a smaller aperture than our light curves and that the galaxy contribution was measured from much higher resolution images than we used here. There may therefore be a difference in the amount of host galaxy flux subtracted, so the fractional variability in this band should not be compared directly with the other light curves.  The X-ray and optical light curves are all highly variable. In agreement with the finding of \citet{shemmer03} that the optical continuum variations are less than the \mbox{X-rays} on time-scales of months, we find that the same is true on time-scales of years. The optical continuum (all light curves, including the 5100\AA$\,$ light curve) is less variable than the X-rays, even after correcting for host galaxy starlight in the aperture.

In the X-ray reprocessing scenario, the fast variations of the X-ray light curve are smoothed out by the disc, so a smaller fractional variation of the optical light curves is expected. Our results are broadly consistent with this expectation, showing a decrease in  $F_{\mbox{\tiny var}}$ towards longer wavelengths. However, due to the uncertainty in the galaxy subtraction, the error on $F_{\mbox{\tiny var}}$ is large, and the decreasing trend is not strong. The observation that the variability of the optical bands is less than that of the X-rays is nevertheless robust. We note that the variability of the $B$ band light curve is lower than the other optical bands and does not follow the decreasing trend. The mean flux of the galaxy-subtracted $B$ band light curve is also slightly higher than that of the $V$ band, which is contrary to expectation, assuming the emission in both bands originate in an accretion disc. This suggests that the low fractional variability in the $B$ band may be the result of underestimating the galaxy contribution in this band.  Increasing the galaxy contribution by 20\%, increases $F_{\mbox{\tiny var}}$ to $15.5\pm1.3$\%.  The larger error may be attributed to the fact that the galaxy spectrum rises steeply in the $B$ band (see Fig.~\ref{fig:galspectrum}), and is likely to be very different from the spectrum of the comparison star used in the differential photometry. The lower variability of the optical with respect to the X-ray variations, suggests that reprocessing of \mbox{X-rays} may account for the optical variability in this source. We return to this possibility in Section~\ref{sec:repr}.

%-----------------

\begin{table}
   \caption{Estimated host galaxy flux inside the 15\arcsec$\,$ aperture and variability characteristics of all light curves, after subtracting the host galaxy contribution.\label{tab:galflux}}
\centering
\begin{tabular}{c c c c}
      \hline\hline
      Light curve & $f_g$ (mJy) & $F_{\mbox{\tiny var}} (\%)$ & $f_{\mbox{\tiny max}}/f_{\mbox{\tiny min}}$ \\
      \hline
      X-ray     & $-$             & $48.8\pm0.8$ & $89.0\pm26.5$ \\
      $u$       & 1.8             & $25.6\pm1.8$ & $3.92\pm1.36$ \\
      $B$       & 6.3             & $11.7\pm1.5$ & $1.92\pm0.56$ \\       	
      $V$       & 16.0            & $15.4\pm3.1$ & $2.48\pm1.64$ \\
      $R$       & 27.2            & $16.9\pm3.8$ & $2.46\pm1.65$ \\       	
      $R1$      & 31.5            & $13.8\pm3.3$ & $2.02\pm0.92$ \\       	
      \vspace{2mm}
      $I$       & 32.7            & $11.8\pm3.0$ & $1.84\pm0.72$ \\       	
      \vspace{1mm}
      $5100$\AA & 7.2$^{\tiny *}$ & $23.9\pm1.5$ & $2.73\pm0.15$ \\
      \hline
   \end{tabular}
   \begin{flushleft}   
      \small{$^{*}$\citet{bentz08}, in the 5.0\arcsec$\times$7.5\arcsec$\,$ spectroscopic aperture.}\\
   \end{flushleft}   
\end{table}

\section{The optical power spectrum} \label{sec:powerspectra}

Power spectra are a commonly used technique to investigate the aperiodic variability displayed by AGN and X-ray binaries. It describes the variability power present in the light curve (mean squared amplitude) as a function of frequency. Generally, the X-ray Power Spectral Density (PSD) of these systems can be described by a power law $P(\nu) \propto \nu^{-\alpha}$ of slope $\alpha \sim 1$ up to a bend or break frequency $\nu_B$, where the power law slope steepens to $\alpha \sim 2$ \citep[e.g.][]{mchardy04psd4051,summonsthesis}. The corresponding break time-scale scales approximately linearly with the black hole mass, and the scatter in this relationship is largely accounted for by considering the difference in accretion rate between these systems \citep{mchardy06nature}. The break frequency is thus characteristic of the system and is thought to represent a time-scale associated with the inner edge of the disc. 

We use the Monte Carlo technique of \citet{uttley02psresp} and \citet{summonsthesis}
to estimate the parameters of the power spectrum underlying the optical variations. The method subtracts the mean of each of the component light curves and then calculates the discrete Fourier Transform of each part. The resulting PSD is binned logarithmically in bins of width 1.5$\nu$, with $\nu$ the frequency at the start of the bin. 

We do not include the AGN Watch data here, as the slightly different wavelength will affect the normalisation of the power and hence the slope we measure. To extend the frequency range, we also include the $V$ band light curve obtained by \citet{klimek04} as part of their microvariability study. We calculate the variability power on each of the five nights it was observed and bin the resulting PSDs together to give the power in the frequency range $6\times10^{-5} - 5\times10^{-4}$~Hz. The low frequency part of the PSD is calculated from the whole of the $V$ band light curve and the middle range from the most intensively sampled part of the light curve from MJD 54181 to 54252.

\begin{figure}
\rotatebox{270}{\includegraphics[width=6.5cm]{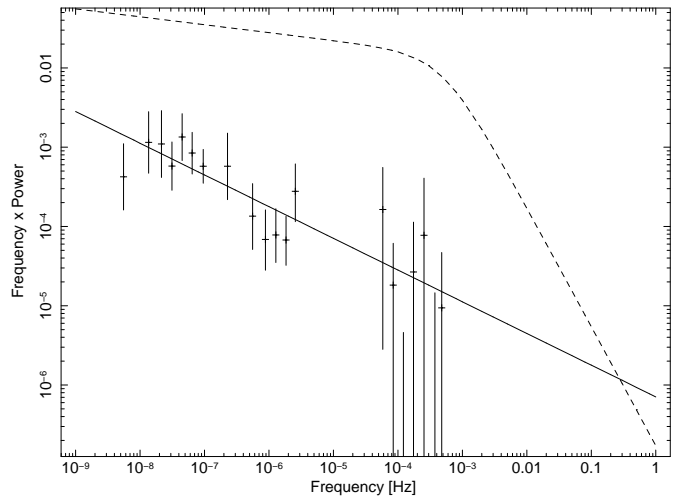}}
\caption{Unbroken power law model (solid line) fitted to the unfolded optical PSD (points with error bars). The slope of the best-fit model is $\alpha=1.4$. The shape of the X-ray power spectrum \citep{summonsthesis} is shown in a dashed line for comparison.}
\label{fig:psdunbroken}
\end{figure}

We start by fitting an unbroken power law
\be P(\nu) = A\nu^{-\alpha} \ee  %
to the power spectrum, allowing the slope $\alpha$ and the normalisation $A$ to vary. The best fit, shown in Figure~\ref{fig:psdunbroken}, is found to have $\alpha=1.4^{+0.6}_{-0.2}$, with an acceptance probability of 79.5 per cent. We also include the shape of the X-ray power spectrum, plotted as a dashed line, for comparison. The X-ray power spectrum has a slope $\alpha_L=1.1^{+0.1}_{-0.4}$ at low frequencies, bending to a slope of $\alpha_H=2.5^{+0.0}_{-0.8}$ and a frequency $\nu_B=5.1^{+4.9}_{-2.6}\times10^{-4}$~Hz \citep[][Summons et al., in preparation]{summonsthesis}. 

The unbroken optical PSD slope is intermediate in value between the low and high frequency X-ray slopes. If the optical PSD has a similar shape to the X-ray PSD, this suggests that a break is contained within the frequency range spanned by the optical data. 
Motivated by the shape of the X-ray PSD, we also fit a bending power law, of the form
\be P(\nu) = \frac{A\nu^{-\alpha_L}}{1+(\nu/\nu_B)^{\alpha_H-\alpha_L}} \ee  %
to the optical PSD. For direct comparison with the X-ray PSD, we fixed the low frequency slope to $\alpha_L=1.1$ and allowed the high frequency slope $\alpha_H$ and bend frequency $\nu_B$ to vary. We find the best fitting parameters to be $\alpha_H=1.5^{+1.0}_{-0.3}$ and bend frequency $\nu_B=1.4^{+*}_{-*}\times10^{-8}$~Hz. The errors are the values for which the acceptance probability drops below 10 per cent and the asterisk is used to indicate an unbounded error. The acceptance probability for this fit is slightly better than for the unbroken case, at 90.2 per cent, but the break frequency is unconstrained by the frequency space sampled by the data.  Allowing all the parameters to simultaneously vary over a large range (i.e. not fixing the value of the low frequency slope), results in the same best-fit values, within the quoted 90\% errorbars.

\begin{figure}
\rotatebox{270}{\includegraphics[width=6.5cm]{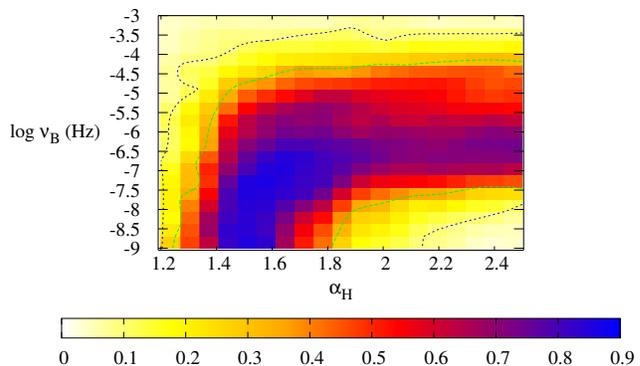}}
%\rotatebox{270}{\includegraphics[width=5.5cm]{figures/contours.ps}}
\caption{Acceptance probability as a function of $\alpha_H$ and $\nu_B$ for the parameter space searched. $\alpha_L$ was fixed at 1.1, for direct comparison with X-ray PSD. The best fit parameters yield an acceptance probability of 0.902. The lines superimposed on the colour plot are the 67\% and 90\% confidence contours.}
\label{fig:psdprob}
\end{figure}

We show in Figure~\ref{fig:psdprob} a colour density plot of the acceptance probability as a function of the parameter space searched, while fixing $\alpha_L=1.1$. It shows that for $\alpha_H=2.5$, as measured from the X-ray spectrum, the best-fit bend frequency is at $\nu_B=4.4^{+*}_{-4.2}\times10^{-7}$~Hz. The fit probability for these parameters is only 74.7 per cent, however, so it can be confidently rejected in favour of the unbroken model.
It is clear from this figure that the best-fit bend frequency in the single-bend model is towards the lowest frequencies (longest time-scales) probed. A single-bend PSD cannot be ruled out for this system, but it would require further monitoring to confirm the existence of a bend at low frequencies. We therefore conclude that, with the current data, the optical power spectrum is best described by an unbroken steep power law model with slope $\alpha=1.4$. 

\section{Cross-correlation analysis} \label{sec:ccf}

%--------------------------------------XRAY-OPTICAL----------------------------------

\subsection{\mbox{X-ray} -- optical correlation}
\label{sec:n4051ccfxrayopt}

In order to test whether there is a measurable delay of one light curve with respect to the other, we calculate their cross-correlation function (CCF), using the interpolation method of  \citet{gaskellsparke86} and \citet{whitepeterson94}. 
We implement the method in two ways: in the first case, we linearly interpolate between adjacent points and resample both light curves to obtain equally spaced points for calculating the correlation. We will refer to this method as the ``equally sampled'' (EQ) method. We note however that for most of the monitoring time, the \mbox{X-ray} light curve is better sampled than the optical light curve. To take advantage of this and maximise the use of real, rather than interpolated data, we use a second implementation of
the method, where we interpolate only the optical light curve and use the observed
fluxes and observation times of the \mbox{X-ray} light curve in the calculation.
We will designate this the ``single interpolation'' (SI) method. The two methods give consistent results throughout, but for completeness we list the cross-correlation results from both methods in Table~\ref{tab:ccfresults}. 

\begin{figure}
\rotatebox{270}{\includegraphics[width=6.5cm]{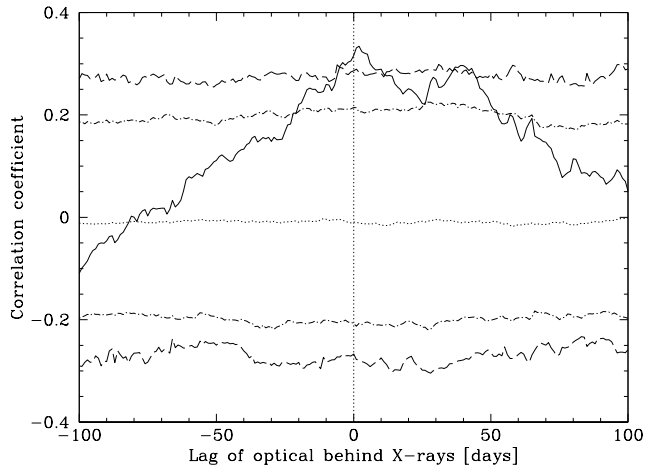}}
\caption{CCF between the \mbox{X-ray} and long term optical light curves, shown in Figure~\ref{fig:xvlc}. The horizontal dotted, dot-dashed and dashed lines are the mean, 95\% and 99\% confidence levels, calculated from 1000 Monte Carlo simulations. The vertical dotted line indicates the zero lag position. Both the peaks are significant at greater than 99\% significance.}
\label{fig:ccfxo}
\end{figure}

To take account of the gaps in the optical light curve when the galaxy is not observable from ground-based observatories, we divide the light curve into segments, defined by these gaps. We calculate the correlation between each segment and its corresponding segment of the \mbox{X-ray} light curve and average the resulting CCFs, weighting them by the length (in time) of each segment. The final CCF, calculated in this way, is shown in Figure~\ref{fig:ccfxo}. It shows a clear peak at $1.2^{+1.0}_{-0.3}$ days, and another at  $38.9^{+2.7}_{-8.4}$ days. 
The errors are calculated using a simple bootstrap method. We select at random two-thirds of the points in the \mbox{X-ray} light curve, each along with its closest optical data point, recalculate the CCF and then measure the peak and centroid. We performed 1000 such selections to yield a distribution of peak and centroid values. The error range reported contains 68\% of the measured values about the median. Here, and throughout this paper, a positive lag indicates the longer wavelength band lagging behind the shorter wavelength band.

The segmentation technique also allows us to take account of the small difference in the flux level of the $V$~band and $5100$\AA$\,$ light curves due to different aperture sizes and a possible difference in the galaxy subtraction.
The mean of each segment is calculated individually and subtracted from the segment before the correlation is calculated. The segmentation has the effect of removing long time-scale power in the light curve and allows a more accurate measure of the short time-scale lag between the light curves. Any small differences in the relative calibration of the light curves are therefore eliminated from the calculation.

We perform Monte Carlo simulations to test the significance of these lags in the following way: we generate 1000 random red noise light curves with the same statistical properties as the observed \mbox{X-ray} light curve, based on the method of \citet{timmerkonig} and the \mbox{X-ray} power spectrum parameters of NGC~4051 \citep[][see also this paper, section~\ref{sec:powerspectra}]{summonsthesis}.
Each of these random, uncorrelated light curves is then sampled in the same way as the observed \mbox{X-ray} light curve and  cross-correlated with the real, observed optical light curve, using the same segmentation technique as described above. The resulting CCF is recorded in each case and compared with the CCF of the observed light curves. In this way we can assess the probability of finding spurious correlations, i.e. chance correlations due to the red noise character of the light curves, that are higher than that of the real light curves. We overlay the mean, 95\% and 99\% levels of these simulations on top of the calculated CCF, in Figure~\ref{fig:ccfxo}. Both peaks reach higher than the 99\% line, showing that the correlations are significant at greater than 99\% confidence. 

\begin{figure}
\rotatebox{270}{\includegraphics[width=6.5cm]{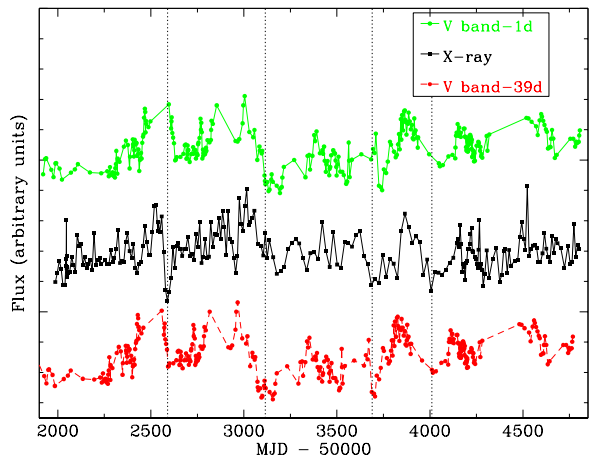}}
\caption{The optical $V$ band light curve shifted back by 1~d (top) and 39~d (bottom), compared to the \mbox{X-ray} light curve (middle). The X-ray light curve was binned in 5~day bins for clarity. The short time-scale correspondence between X-ray and the optical light curve at the top is clear. The vertical lines indicate some of the variations which align better when the optical light curve is shifted back by 39~days, rather than by the main 1~day lag.}
\label{fig:laglcs}
\end{figure}

To illustrate the origin of the two lags in the light curves, we  show in Figure~\ref{fig:laglcs} the optical light curve shifted by 1 (top) and by 39 days (bottom), in comparison to the \mbox{X-ray} light curve (middle). The X-ray light curve was binned in 5-day bins for clarity. Most of the X-ray variations on time-scales of days to weeks show corresponding variations in the optical light curve shifted by 1 day. However, a few of the minima and maxima align better with the 39-day shifted version of the optical light curve. We point out that since both lags are short compared to the total duration of the light curve, only the largest amplitude variations can be clearly seen in Figure~\ref{fig:laglcs}. The most obvious examples of the delayed variations are marked by dotted lines in that figure.

%--------------------------------------COLOUR-LAGS----------------------------------

\subsection{Lags between optical bands}
It is clear from the light curves in Figure~\ref{fig:colourlcs} that a strong correlation exists between the optical bands. Optical interband lags in this source have been measured by \citet{sergeevetal05} and \citet{cackettetal07}, confirming the apparent correlation. We consider here the relationship between the \mbox{X-rays} and each of the optical bands.

\begin{figure}
\rotatebox{0}{\includegraphics[width=9cm]{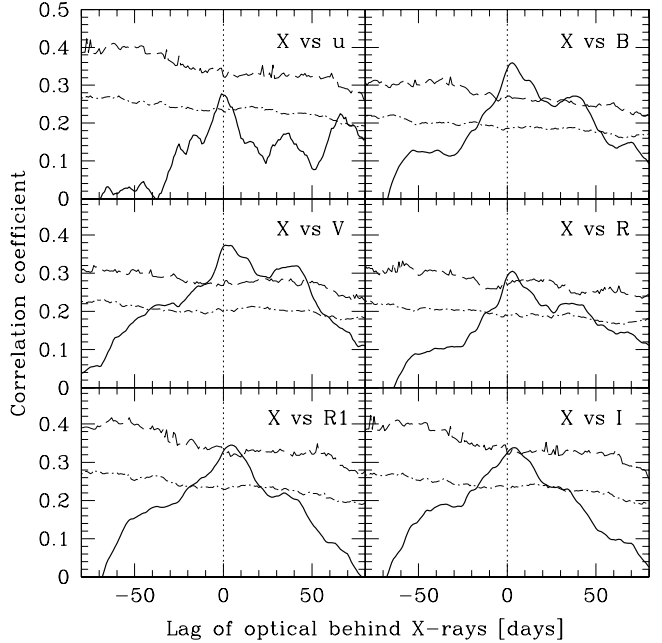}}
\caption{CCF between the \mbox{X-ray} and optical bands as labeled. The vertical dotted line indicates the zero lag position. For simplicity, we only show the 95\% (dot-dashed line) and 99\% (dashed line) confidence levels.}
\label{fig:colourccfs}
\end{figure}

Using the same segmentation technique as before, we calculate the correlation between the \mbox{X-ray} light curve and each of the different optical bands. The CCFs are shown in Figure~\ref{fig:colourccfs} and we summarise in Table~\ref{tab:ccfresults} the value of the peak and centroid of the CCF, calculated with both methods of interpolation. The centroid is calculated as the weighted average of the CCF points around the peak, at a value \mbox{$\geq 85\%$} of the peak correlation coefficient. As before, a positive value of the lag indicates that the variations in the optical band are lagging behind the \mbox{X-ray} variations.  All light curves, except the $u$ band, for which the sampling is much sparser than the other light curves, show a correlation significant at greater than 99\% confidence. The second peak near $\sim40$~days is also seen, though not always at a significant level. We note however, that the strength of the second peak, {\em relative} to the peak at $\sim0$~days, remains approximately constant. 
The $V$ band light curve used in this part of the analysis, consists of only those observations which overlap in time with the other colour light curves. From this segment of the data we measure a slightly shorter lag than from the full light curve, but the results are consistent within the $1\sigma$ errors. We also note that the lags measured from the central CCF peak are well within the $\sim6$~light day estimate of the location of the H$\beta$ broad line region (BLR) in this source \citep{peterson00, kaspi00, peterson04}, indicating that most of the variable optical emission originates in a structure smaller than the BLR.

In order to assess the influence of the delayed line emission on the measured lag, we first estimate the line contribution to the flux measured through the broad band filters. Using an average optical spectrum from the AGN Watch program, convolved with the filter response, we find a 3\% line contribution to the $V$ band flux and a 4\% contribution to the $B$ band flux. 
We then generate correlated X-ray and optical light curves, using the method described in Section~\ref{sec:n4051ccfxrayopt}, to represent the continuum emission. The line flux is represented as a smoothed version of the X-ray light curve, delayed by 6 days, as would be appropriate for a thin spherical shell of radius 6 light-days. The delayed line flux is added to the simulated, undelayed optical light curve in the ratios measured from the spectrum. 1000 such simulations yield a peak lag of $\tau_{\mbox{\tiny peak}}=0.0\pm0.1$ days and a centroid lag of $\tau_{\mbox{\tiny cent}}=0.1\pm0.6$ days, which is well within the errors on the lag meaured in the real light curve. We therefore conclude that the line emission contributes too small a fraction of the total $V$ band emission to noticably affect the measured lag.

%--------------------------------------TABLE:-LAG-RESULTS----------------------------------

\begin{table}
      \caption{Cross-correlation lags (in days). The centroid is the weighted
                average of the CCF values $\geq$ 85\% of the peak correlation coefficient. 		\label{tab:ccfresults}}
\begin{tabular}{c c c c c}
      \hline\hline
      CCF & Peak & Centroid & Peak & Centroid \\
      X-ray vs. & (EQ) & (EQ) & (SI) & (SI) \\
      \hline
      Optical$^{\tiny \P}$&  &  &   & \\
      \vspace{1mm}
      {\em first peak} & $2.0^{+1.2}_{-0.6}$  & $2.4^{+0.9}_{-1.6}$  & $1.2^{+1.0}_{-0.3}$  & $1.8^{+1.3}_{-2.0}$\\
      \vspace{3mm}
      {\em second peak}& $35.2^{+6.6}_{-5.1}$ & $39.4^{+2.2}_{-2.1}$ & $38.9^{+2.7}_{-8.4}$ & $38.7^{+1.3}_{-1.0}$\\
      \vspace{1mm}
	$u$ & $0.3^{+1.5}_{-1.3}$ & $0.5^{+1.4}_{-1.6} $ & $-0.2^{+1.6}_{-1.1}$ & $0.1^{+2.0}_{-1.5}$\\
      \vspace{1mm}
	$B$ & $2.3^{+1.5}_{-0.7}$ & $2.6^{+0.5}_{-0.6} $ & $1.5^{+0.8}_{-0.6}$ & $2.0^{+0.8}_{-1.1}$\\
      \vspace{1mm}
	$V$ & $1.1^{+0.5}_{-0.3}$ & $2.2^{+0.3}_{-0.4} $ & $0.6^{+0.4}_{-0.6}$ & $1.3^{+0.8}_{-1.0}$\\
      \vspace{1mm}
	$R$ & $3.1^{+1.0}_{-1.6}$ & $2.8^{+0.6}_{-0.7} $ & $1.5^{+1.6}_{-0.6}$ & $2.0^{+0.9}_{-1.1}$\\
      \vspace{1mm}
	$R1$ & $4.8^{+2.1}_{-3.3}$ & $2.3^{+1.3}_{-1.7} $ & $1.9^{+3.1}_{-1.0}$ & $2.0^{+1.3}_{-2.2}$\\
      \vspace{1mm}
	$I$ & $4.9^{+1.9}_{-2.4}$ & $2.0^{+1.2}_{-1.9} $ & $1.7^{+3.3}_{-0.8}$ & $1.7^{+1.0}_{-1.7}$\\
      \hline
   \end{tabular}
   \begin{flushleft}   
      \small{$^{\P}$This is the light curve shown in Fig.~\ref{fig:xvlc}, including both the $V$ band data and 5100\AA$\,$ continuum AGN Watch data.}\\
   \end{flushleft}   
\end{table}

%--------------------------------------REPROCESSING----------------------------------
\section{Reprocessing} \label{sec:repr}

The short, but measurable delays between the \mbox{X-rays} and the optical bands, as well as the small amplitude of the optical variations, suggest that reprocessing of X-rays plays a role in producing the optical variations. To test this hypothesis, we use the observed \mbox{X-ray} light curve to construct a model reprocessed optical light curve, which we can compare to the observed optical light curve. 

The model, described in detail by \citet{reprocessing}, assumes that the disc temperature is determined by the combined effects of viscous dissipation in the disc and variable heating by an \mbox{X-ray} source at height $h_x$ above the disc and on its axis of symmetry. The temperature of the disc is then given by % 
\be T(R,t)\!=\!\!\left[\frac{3\dot{M}c^2}{8\pi\sigma R_g^2} \frac{1}{R^3}\!\!  \left(\!\!1\!-\!\!\sqrt\frac{R_{\mbox{\tiny{in}}}}{R}\!\right) \!+\!  \frac{(1\!-\!\mathcal{A})L_x(t)}{4\pi\sigma (h_x^2\!+\!R^2)} \cos\theta\right]^{\frac{1}{4}} \label{eq:disctemp} \ee %
assuming that each annulus of the disc emits as a blackbody. Here $\dot{M}$ is the disc accretion rate in kg~s$^{-1}$, $R$ the distance of the disc surface element from the centre, $R_{\mbox{{\tiny{in}}}}$ the location of the inner edge of the accretion disc, $L_x(t)$ the variable \mbox{X-ray} luminosity, $\mathcal{A}$ is the average disc albedo and $\theta$ the angle between the disc surface and the direction of the \mbox{X-ray} source. $R$, $R_{\mbox{{\tiny{in}}}}$ and $h_x$ are measured in units of the gravitational radius $R_g=GM/c^2$, where $M$ is the mass of the black hole in kg. The delay between the variations of different wavebands is then due to the difference in light travel time to regions of the disc of different temperature.

\begin{figure}
\begin{center}
\rotatebox{0}{\includegraphics[width=6cm]{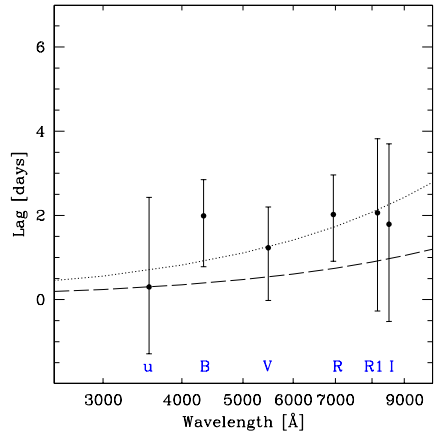}}
\end{center}
\caption{Centroid lags measured from the CCFs between the \mbox{X-rays} and optical colours. The dashed line shows the predicted lags from an irradiated accretion disc model for parameters appropriate to NGC~4051 (see text). The dotted line is a simple $\lambda^{4/3}$ model, fitted to the data by an arbitrary scaling constant.}
\label{fig:colourlags}
\end{figure}

We show in Figure~\ref{fig:colourlags} the lags predicted by this model, plotted as a dashed line, assuming $M=1.91\times10^6$\msun \citep{peterson04}, $\dot{M}=0.15\dot{M}_{\mbox{\tiny{Edd}}}$ \citep{woourry02}, $R{\mbox{{\tiny{in}}}}=6$, $h_x=10$, $\mathcal{A}=0.3$ and a constant X-ray luminosity of $L_x=3\times10^{42}$ erg s$^{-1}$ \citep{smithvaughan06}. The points with error bars are the measured centroid lags (of the light curves shown in Figure~\ref{fig:colourlcs} versus the X-rays) as a function of wavelength. With the above fixed parameters, the irradiated disc model (Eqn.~\ref{eq:disctemp}) is not a particularly
good fit. Although, as we show next, a much better fit can be obtained if we alter the parameters (e.g. decrease the mass or increase the accretion rate).

Note that both terms in Eqn.~\ref{eq:disctemp} scale as $R^{-3/4}$ for large radii. `Large' here implies $R\gg R_{\mbox{\tiny{in}}}$ and $R\gg h_x$, since we can write $\cos\theta = h_x/(h_x^2+R^2)^{1/2}$. Together with the assumption that the peak of the emission scales as $T\propto\lambda^{-1}$, and writing the delay between different bands as $\tau=R/c$, we obtain the well-known $\tau\propto\lambda^{4/3}$ relationship commonly quoted in the context of reprocessing models.
(Of course, where the large radius assumptions break down the relationship between the lag and the emitted wavelength is somewhat modified; see also \citealt{gaskell08}).
We show in Figure~\ref{fig:colourlags} (dotted line) that a simple $\lambda^{4/3}$ model, fitted to the measured lags via an arbitrary scaling constant, fits the data well. The factor of $\sim$2 difference between the two curves can easily be accounted for within the uncertainties on the model parameters.

%-------------------------------------------------

To construct a model light curve, the emitted flux as a function of time may be computed by integrating the emission over all light travel time delays $\tau$ to different parts of the disc, and throughout all disc radii,%
\be f_\lambda(t) = \int_{R_{\mbox{\tiny{in}}}}^{R_{\mbox{\tiny{out}}}}dR \int_{\tau_{1}}^{\tau_{2}}A(R,\tau)B_\lambda\bigl( T(R,t-\tau)\bigr) \; d\tau \ee %
where $T(R,t)$ is given by Eqn.~\ref{eq:disctemp}. $A$ is an area function obtained by considering the effect of the disc inclination on the distance between the X-ray source and the disc \citep[see][]{BKO2000}, and $B_\lambda(T)$ is the Planck function. $R_{\mbox{\tiny{out}}}$ is a large value, appropriate for the outer edge of the disc. 

We allowed the inner disc radius $R_{\mbox{\tiny{in}}}$, the accretion rate $\dot{m}=\dot{M}/\dot{M}_{\mbox{\tiny{Edd}}}$, height of the X-ray source $h_x$ and disc inclination $\theta$ to vary, to find the parameters best describing the observed optical light curve. The mass was fixed at the reverberation-mapped value, $1.91\times10^6$\msun \citep{peterson04}. 

\begin{figure}
\rotatebox{270}{\includegraphics[width=6.5cm]{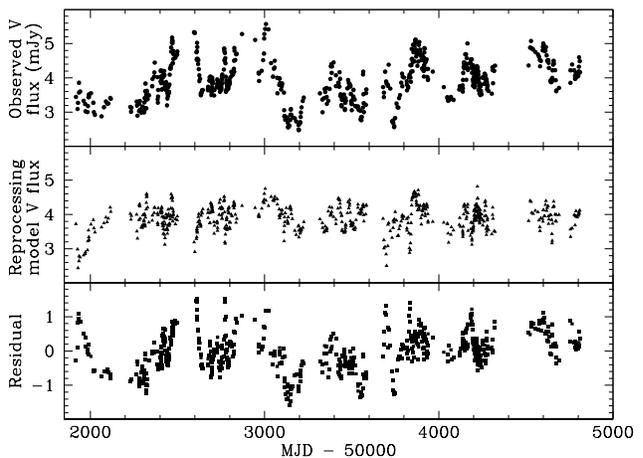}}
\caption{{\em Top panel:} Galaxy-subtracted $V$ band light curve. {\em Middle panel:} Best-fit model reprocessed $V$ band light curve, using the observed X-ray light curve as input. {\em Bottom panel:} Residual variations in the optical light curve after the reprocessed component has been removed.}
\label{fig:difflc}
\end{figure}

Figure~\ref{fig:difflc} shows the observed optical light curve compared to a typical model reprocessed light curve. Due to the short light-travel time to the optical emitting region, the rapid variability of the X-rays is preserved in the model light curves and we find that a range of input parameter values produce nearly identical model light curves. As expected, the model reprocessed light curve is highly correlated with the X-ray light curve, trailing it by $0.4-0.8$~days, depending on the input parameters. 

This is a very simple model, which assumes the X-rays to be emitted by a compact source located on the symmetry axis of the system, at a height $h_x$ above the disc surface. In reality, it is more likely that the X-rays come from an extended corona. The compact source model used here may be viewed as a first order approximation of the centroid of such a corona. We believe that this simplified model is sufficient in this case, as $h_x$ is much smaller than the radius where most of the optical emission is coming from. The break in the X-ray power spectrum, which is thought to be associated with the inner edge of the disc, corresponds to the dynamical time-scale at $35R_g$, or the viscous time-scale of a thick disc ($\alpha=0.1, H/R\sim0.5$) at $3R_g$, 
so the bulk of the \mbox{X-rays} are probably emitted within a radius smaller than, or comparable to this. Using Comptonization models \citet{uttley00} estimate the size of the emission region to be $< 20R_g$. Integrating the blackbody emission through the disc, we find that 95\% of the $V$~band emission originates from outside a radius of 400$R_g$. %
From this distance, even an extended  corona would resemble a point source, making a more detailed model of the X-ray source unnecessary.

\begin{figure}
\rotatebox{270}{\includegraphics[width=6.5cm]{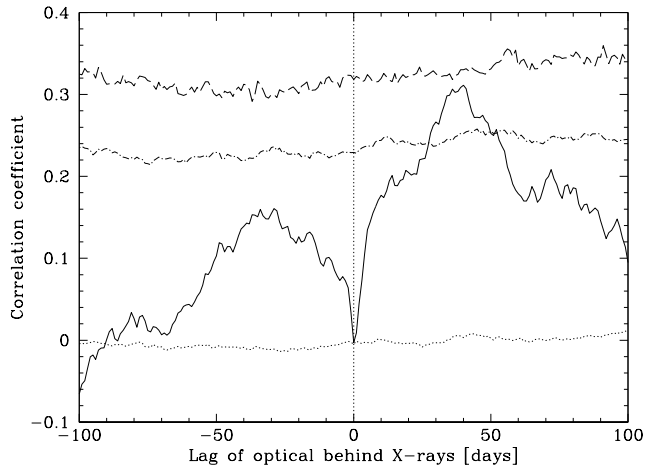}}
\caption{Cross correlation of the reprocessing-subtracted residual light curve with the X-rays. The $\sim0$~day peak is removed entirely from the CCF, leaving only the longer time-scale correlation peak. The horizontal lines are the 50\% (dotted), 95\% (dot-dashed) and 99\% (dashed) significance levels. The peak is significant at 98.4\%.}
\label{fig:diffccf}
\end{figure}

Reprocessing by a standard, thin disc appears to describe the observed lag and the short time-scale ($\sim$days) variations of the optical light curve well, but it fails to simultaneously reproduce the larger amplitude flares on time-scales of $\sim$months. % 
Closer inspection of the light curves in Figure~\ref{fig:xvlc} confirms that it is these larger amplitude fluctuations which lag the X-rays by $\sim40$~days (see e.g. MJD $\sim$ 50750, 52500 and 53700 for obvious examples).
Attempting to isolate these variations, we show in the bottom panel of Figure~\ref{fig:difflc}, the residual of  subtracting the reprocessing model light curve from the observed light curve. Cross correlation of the X-rays with this ``residual curve'' yields the CCF shown in Figure~\ref{fig:diffccf}.
As the model light curve is very well correlated with the X-rays at near 0~days, subtracting it removes this correlation peak entirely from the CCF (the correlation at this lag drops to zero).  Isolating the second peak this way, allows us to make a more precise measurement of the lag it corresponds to. The peak is found to be at $38.9^{+1.2}_{-3.3}$~days, and the 85\% centroid at $38.5^{+2.5}_{-2.2}$~days. We calculate the significance of the correlation as before, by cross-correlating random red noise light curves with the difference light curve. It is found to be significant at 98.4\% confidence. It is therefore clear that reprocessing by a flat disc can only account for the short lags in the light curve and that the second peak must be produced by optical emission from a structure further out.

\section{Discussion} \label{sec:discussion}

The detection of two significant peaks in the X-ray/optical correlation function suggest that the optical variability originates from more than one location in the system, or through more than one process. In the previous section we discussed the correlation at short lags in terms of reprocessing by an accretion disc. We showed that this model can only reproduce the short time-scale variations in the light curve, and that the second peak must be the result of optical emission coming from beyond the disc and BLR.

In this section we investigate the origin of the second peak in the correlation function. The lag is found to be consistent with the light travel time to the inner radius of the dust torus in this source, so we investigate the possibility of optical emission originating from the torus. Firstly we will consider reprocessing of X-rays by the dust, and secondly the dust reflecting optical light from the accretion disc.

%--------------------------------------TORUS-IR----------------------------------

\subsection{Second peak in the CCF -- optical emission from the torus?} \label{sec:ir}

The standard model of AGN holds that the disc and BLR is surrounded by a large molecular torus. It is now widely accepted that the infrared emission from AGN is the result of reprocessing of high energy radiation by hot dust in the torus \citep[e.g.][]{glass04}. We investigate here whether it is possible that the second peak detected in the correlation function could be the result of optical emission coming from the torus.

\subsubsection{Distance to the torus}

The inner radius of the dust distribution is set by the dust sublimation radius, inside which the flux from the central source will destroy the dust particles. Detailed radiative transfer calculations \citep{nenkova08a} places this radius at \be r_{\mbox{\tiny{subl}}}=0.4\left(\frac{L}{10^{45}\,\mbox{erg s}^{-1}}\right)^{0.5}\left(\frac{T}{1500\,\mbox{K}}\right)^{-2.6}\;\mbox{pc}. \label{eq:rsubl} \ee %
where $T$ is the dust temperature and $L$ the UV-optical luminosity.

Where adequate data exist, the optical to near-infrared delays show reasonable agreement with this model. Using both their own monitoring data as well as results from literature, \citet{oknyanskijhorne01}, \citet{minezaki04} and \citet{suganumaetal06} show that the optical-to-infrared lag is proportional to $L^{0.5}$ over a range of luminosities.

The exact temperature at which the dust in the torus sublimates is not well known, as it depends on the composition of the dust. The dust is generally assumed to be a mixture of graphite and silicate particles, with sublimation temperatures in the range 1500--2000~K. The emission from hot dust at the inner edge of the torus is therefore expected to peak in the $K$ band. As graphite can survive higher temperatures, it is assumed that the dust at the inner edge will mostly consist of graphite grains.

\citet{gaskell07} pointed out that at such high temperatures, there must be a small, yet significant amount of optical emission coming from the torus as well, as part of the Wien tail of the $\sim$1500--2000~K emission. He proposed that the increase in time delay measured between the optical bands could be the result of contamination by the additional optical emission coming from large radii.  Optical emission coming from the torus will of course also be delayed with respect to the X-rays, by the same amount as the infrared emission. This will cause an asymmetry in the X-ray--optical correlation function, or, if the signal is strong enough, another peak may be seen, corresponding to the light travel time to the inner edge of the torus. The dust recombination time-scale is expected to be long compared to the sublimation time-scale \citep{koshida09}, so that the inner edge of the torus is determined by the largest of the X-ray/UV flares. Any variations we see in the emitted flux is then probably due to temperature variations of the dust (below the sublimation temperature), rather than a change of the torus inner radius. 

The $H-K$ colour measured for NGC~4051 by \citet{suganumaetal06}, corresponds to a blackbody temperature of approximately 1600~K, so we adopt this value as the temperature of the dust in this source. We scale the ionizing X-ray-UV-optical luminosity ($\sim$13.6~eV--13.6~keV), estimated by \citet{ogle04}, to a distance of 15.2~Mpc, giving $L_{\mbox{\tiny{ion}}}=1.1\times10^{43}$~erg~s$^{-1}$. The sublimation radius (Eqn.~\ref{eq:rsubl}) of NGC~4051 is then $r_{\mbox{\tiny{subl}}}=42.2$~light-days. 
The position of the second peak in our correlation function agrees well with this value. We note, however, that the $V$-to-$K$ lag found by \citet{suganumaetal06} is only $\sim20$~light-days. This may be expected if the disc $V$ band emission is contaminated by optical emission from large radii (such as the torus), as this will shift the peak of the $V$ vs. $K$ CCF to shorter lags. Several studies \citep{peterson00,kaspi00,peterson04} measure the location of the BLR to be at $\sim6$~light-days, placing even the smallest estimates of the torus well outside the disc and BLR.

\subsubsection{Energetics}

Detailed models of AGN dust torii (e.g. \citealt{nenkova08a}a, \citealt{nenkova08b}b) take account of the fact that the torus probably consist of clumpy material rather than a smooth distribution of dust. One effect the clumps has is to cause a steep radial temperature gradient in the torus, as optically thick clouds closer to the source can shield those further away from direct irradiation. It also takes into account the diffuse heating of clumps by each other. Such a detailed analysis is beyond the scope of this paper, but as we are mainly interested in the hottest dust here, a simplified approach may be sufficient to determine whether it is energetically possible to get the 1--2~mJy $V$ band flux required to explain the large amplitude optical fluctuations, from the torus. 

A blackbody at 1600~K emits roughly 16\% of its total emission in the $K$ band, and only 0.04\% in the Wien tail at $V$ band wavelengths. The fraction of $V$ band emission decreases very rapidly as the dust temperature decreases, so we do not expect to find a significant contribution to the optical from further into the torus. We will assume that the optical emission is identified with the inner edge of the dust distribution only, where the hottest dust is located. For simplicity, we also assume a sharp boundary, although the transition is most likely a gradual one, given the difference in sublimation temperatures of different dust particles. 
We follow the prescription of \citet{barvainis87} to calculate the luminosity emitted by the dust. We assume a grain radius of $a=0.05\mu$m and infrared ($K$ band) absorption efficiency $Q_{\mbox{\tiny{abs}}}=0.058$. The luminosity of a single grain is given by %
$L_{\nu}^{\mbox{\tiny{gr}}} = 4\pi a^2\,\pi Q_{\nu}\,B_{\nu}(T)$~erg~s$^{-1}$~Hz$^{-1}$, where $B_\nu$ is the Planck spectrum of a grain at temperature $T$. The $K$ band spectral luminosity of a graphite grain at $T=1600$~K is then $L_K^{\mbox{\tiny{gr}}} = 3.6\times10^{-17}$~erg~s$^{-1}$~Hz$^{-1}$. The average galaxy-subtracted $K$ band flux measured by \citet{suganumaetal06} is 31~mJy, which corresponds to a total $K$ band luminosity of $L_K = 8.6\times10^{27}$~erg~s$^{-1}$~Hz$^{-1}$. Hence the total number of hot dust grains is $N\sim2.4\times10^{44}$.  %
For UV-optical radiation, the absorption efficiency is $Q_{\mbox{\tiny{abs}}}\sim1$ for grains of this size \citep{wickramasinghe74}. The optical emission of a grain at 1600~K is  $L_V^{\mbox{\tiny{gr}}} = 1.9\times10^{-19}$~erg~s$^{-1}$~Hz$^{-1}$. If we assume the same number of grains to be emitting this optical emission (i.e. only the hottest dust) then the total $V$ band luminosity we may expect from the torus is $L_V = 4.4\times10^{25}$~erg~s$^{-1}$~Hz$^{-1}$. This translates to an observed flux of only 0.16~mJy, which is an order of magnitude smaller than required to explain the additional variability in the observed $V$ band light curve.

One of the main uncertainties in this model is the dust composition, as there is no guarantee that dust in the vicinity of the active nucleus will be the same as dust found in molecular clouds. The temperature of the grains is determined by the balance between UV absorption and infrared emission, which vary with grain size and type. This, in turn, affects the sublimation temperature. If we allow the dust temperature to increase to 1800~K (which is still within range of the sublimation temperature of graphite), the emitted luminosity per grain increases to  $L_V^{\mbox{\tiny{gr}}} = 1.15\times10^{-18}$~erg~s$^{-1}$~Hz$^{-1}$, and the additional flux observed in the $V$ band to 1.1~mJy. Such an increase in the temperature the dust particles can sustain will of course move the sublimation radius inwards as well, to a distance of $r_{\mbox{\tiny{subl}}}=31.1$~light-days. This is within the $1\sigma$ error of the peak measured on the X-ray--optical CCF (Figure~\ref{fig:ccfxo}), and within $2\sigma$ of the peak measured from the CCF calculated using residual light curve (Figure~\ref{fig:diffccf}). The uncertainty on $L_{\mbox{\tiny{ion}}}$ would also allow a factor of $\sim$2 change in the sublimation radius.
A similar analysis for the other optical bands, using $T=1800$~K and the appropriate absorption efficiencies, yields $u=0.004$, $B=0.08$, $R=5.8$, $R1=13.2$ and $I=16.6$~mJy. For wavelengths $R$ and longer, the reprocessing model therefore predicts a higher flux to come from the torus alone than the total galaxy-subtracted flux we measure. The measured fluxes in these bands must however include a significant contribution from the disc as well, as their correlation functions with the X-rays (Fig.~\ref{fig:colourccfs}) all display a significant peak at short lags. 

For increasing wavelength, the torus reprocessing model naturally predicts an increase in the amount of reprocessed flux originating in the torus, as the redder bands include a larger fraction of the total flux emitted by the dust. Our data do not appear to support such an increase, so it is unlikely that all the additional optical flux results from dust reprocessing. We nevertheless cannot exclude the possibility that some of the optical emission in this source is originating in the dust torus.

\subsubsection{Optical reflection}

In this section we consider the reflection of optical light by the torus.  Apart from the illumination by the central source, the torus will also receive UV and optical light from the accretion disc. The dust absorbs energy at these wavelengths very efficiently, so   some of this reprocessed optical emission will be absorbed by the dust and contribute to the variable heating of the torus, as discussed in the previous section. Depending on the solid angle subtended by the torus at the optically-emitting part of the disc, some of the optical emission may also be elastically scattered and reflected by the torus. The efficiency of reflection is strongly geometry dependent, so we will limit our discussion here to a qualitative description of the main observable results expected. Reflection from different geometries have been considered by \citet{goosmanngaskell_stokes} in the development of their Monte Carlo polarization code {\sc stokes}. 

Most of the UV/optical illumination will be relatively steady, or will vary on the very long (years) viscous time-scales on which the intrinsic emission from the disc will vary. However, as we have already shown, a small fraction of the optical illumination, resulting from reprocessing of X-rays by the accretion disc, will follow the pattern of the X-ray variations. As the optically-emitting region is about a light-day away from the central X-ray source, the torus will see optical emission which broadly follows the X-ray emission,
but which is smoothed on a time-scale of a day or two relative to the X-ray variability pattern. 

The component of the optical emission illuminating the torus which has a phase relationship with the X-rays (so that it causes a peak in the CCF), must have arisen from reprocessing of X-rays in the disc. The amount of scattered light depends on the solid angle the torus subtends at the optically emitting part of the disc, 
so for large solid angles, the elastically scattered optical photons may produce a noticeable contribution to the optical variations, and will lag the X-rays by about 40~days. Dust models by \citet{draine03} suggest that the scattering albedo (scattering cross section as a fraction of the total extinction cross section) of Milky Way type dust may be as high as 0.5--0.8, for different dust compositions. Dust in AGN probably consists of larger grains (smaller grains are evaporated by the intense radiation field), so the albedo is uncertain, but it may still be high enough for the scattered light to contribute significantly to the observed optical radiation. 

Reflection models are supported by spectropolarimetric studies which are used to detect the broad line region in Seyfert~2 galaxies, such as NGC~1068, through polarized light. In these systems, the continuum source and BLR clouds are hidden from our view by the molecular torus, but their emission is scattered into our line of sight by warm electrons and dust lying outside the torus opening, polarizing the light. In Seyfert~1 galaxies, however, the optical polarization position angle is generally observed to be parallel to the radio axis (which is assumed to be the system axis), suggesting that the scattering medium has an equatorial geometry, perhaps a ring of material co-planar with with accretion disc \citep{smith02,smith04}, or the inner edge of the torus \citep{cohenmartel02}. The optical continuum polarization in NGC~4051, as measured by \citet{smith02}, is low however. They find  0.5 per cent of the optical continuum at 7000\AA$\,$ to be polarized, amounting to a mere $\sim0.05$~mJy. It is also not clear to what extent this is intrinsic to the source or whether the measurement is contaminated by foreground interstellar polarization. If intrinsic, the position angle of the polarization is approximately parallel to the radio axis, suggesting that the scattering occurs in the equatorial region. Unfortunately their polarization observation was made during a time for which we do not have optical monitoring data available, so we cannot comment on the general optical behaviour of the source at that time.

There is also evidence for diffuse continuum emission and reprocessing by the BLR, as well as reflection by the broad line clouds \citep{koristaferland98,koristagoad01,arevalo09} in some Seyfert galaxies. Though, as mentioned before, any lags associated with the BLR ($\sim$6 days for NGC~4051) are likely to be much shorter than those associated with the torus.

As an aside, we note that the high energy emission may also be reflected by the torus. X-ray spectra show evidence of X-ray reflection by high density gas in the form of the ``Compton bump'' and the Fe K$\alpha$ fluorescence line at 6.4~keV. This line is thought to originate in the accretion disc, and is therefore expected to be broadened by relativistic Doppler motions. Broad Fe lines like this are seen in X-ray spectra, but the line is found to have a narrow component as well, which is associated with reflection by distant, optically thick material such as the molecular torus \citep[e.g.][]{nandra07}.

Neither the reprocessing nor the scattering model can on its own fully account for the optical flux causing the second peak in the correlation function, but it is possible that either or both may contribute to the optical emission observed from the system.
Overall we conclude that there are many possible sources of optical emission in the inner parts of AGN but all those associated with the torus, albeit reprocessing or simple reflection, will give more or less the same lag. Our current data is not suited to constrain these models further, so a more detailed discussion is beyond the scope of this paper.

%-----------------------------------------------------------------------------------------------%

\subsection{The complex optical variability in NGC~4051}

\citet{shemmer03} find that part of the optical emission leads the X-ray variations by approximately 2~days, contrary to what we find here (see Table~\ref{tab:ccfresults}).
As was shown in previous sections, there is more than one effect contributing to the optical variations in this source. Variations on time-scales longer than, or similar to, their 3 month monitoring period may cause asymmetries in the correlation function and hence a centroid at negative lags. \citet{shemmer03} interpret their result as a combination of reprocessing and longer time-scale variations, similar to what we find over much longer time-scales. We note that the peak of their cross-correlation function is consistent with the peak lag we find here, $\sim1$~day.

In red noise time series, such as the X-ray and optical light curves used here, peaks and troughs may sometimes line up even if the light curves are not really correlated, producing spurious peaks in the CCF. 
We believe that the second peak we measure in the CCF here, is not simply due to statistical fluctuations like these, as it is present in most of the individual segments used to calculate the final CCFs. If this was an effect only due to a random correlation in a red noise process, we would expect to see it in, e.g. only one of the segments' CCFs.  We note that the peak does shift between $\sim30-50$~days in the CCFs of the different segments. The probability that the peak is due to such statistical fluctuations was assessed through Monte Carlo simulations in Section~\ref{sec:ccf}, and was found to be 1.6\%.

So far we have interpreted the second peak in the CCF only in terms of a light crossing time-scale. Other time-scales of interest are the dynamical (Kepler) time-scale, and the viscous time-scale, on which we expect accretion rate fluctuations to propagate through the disc. A lag of the optical variations behind the X-rays, as seen here, would require such fluctuations to be propagating outwards through the disc for this time-scale to be relevant. A viscous time-scale of 40 days corresponds to $R=50R_g$ in a moderately thin, optically thick disc ($H/R\sim 0.1$, $\alpha_{\mbox{\tiny{disc}}}=0.1$), or to $R=440R_g$, in a geometrically thick disc ($H/R\sim 0.5$, $\alpha_{\mbox{\tiny{disc}}}=0.1$). $R=50R_g$ lies well within the Far- to Extreme UV emitting part of the disc; less than 0.5\% of the optical emission come from within this radius.  $R=440R_g$ contains approximately 6\% of the optical emission. Enhanced emission due to viscous instability in this region could therefore only cause a $\sim6$\% variation in the light curve. This translates to a flux variation of 0.2~mJy, which is not enough to explain the variations seen in the light curve.  
As an orbital time-scale, 40~days corresponds to the radius which contains 69\% of the $V$ band emission, but is is difficult to associate an orbital time-scale with a lag between the emission in different wavebands, as we see here.

If this is indeed optical emission from the torus, why has this not been seen in other sources?  NGC~4051 is a low mass, low luminosity system, which means that it is physically smaller than higher mass AGN. Lags between the optical and near-infrared suggest that dust sublimation radii in most sources are several tens to hundreds of days, but here it is only $\sim$20--40 days. The optical emission from the torus will be a very weak signal, so long enough light curves are necessary to include several cycles of the correlated variations to produce a significant peak in the CCF. We note that calculating the CCF using only half of the $V$ band data, decreases the significance of the second peak in Figure~\ref{fig:ccfxo} to 96.3\%, or to below 95\% if only a quarter of the light curve is used. 

More importantly, in low mass systems the $V$ band emission originating in the torus is probably a larger fraction of the total $V$ band emission emitted by the system, than it is in more massive systems. This is due to the variation in disc temperature with black hole mass. For example, for a fixed accretion rate of $\dot{m}=0.10$, a standard disc \citep[e.g.][]{fkr} around a $10^8$\msun black hole emits 1.9\% of its total luminosity at $V$ band wavelengths, but for a $10^6$\msun black hole the disc is hotter, the bulk of the energy is emitted at bluer wavelengths and the fraction of the total emission emitted at $V$ band wavelengths decreases to 0.4\%. On the contrary, the torus in all sources is expected to be at approximately the same temperature (as its inner radius and temperature is determined by dust sublimation) and hence it will contribute a constant fraction of the $V$ band emission. Comparing this constant fraction with the $V$ band emission originating in the disc, the torus contributes a larger fraction to the total $V$ band emission in the $10^6$\msun source than it does in the $10^8$\msun source.  In more massive systems the $V$ band emission is therefore dominated by the disc, so the weak optical emission from the torus  will be very hard to isolate.

Long-term multicolour optical and infrared monitoring of NGC~4051 is needed to reliably measure the inner radius of the dust torus and the possible contribution of the disc to the near-infrared \citep[e.g.][]{tomita06}. Detailed modeling of the torus emission, taking into account both the geometry of the system as well as the dust albedo and emission properties, is also necessary to properly assess the contribution this could have to the optical emission observed. 
\section{Summary} \label{sec:conclusion}

We have presented a 12.6~year 3--10~keV X-ray light curve from our ongoing long term X-ray monitoring program on \rxte, along with concurrent optical monitoring light curves. We combined the $V$ band data from four different telescopes with archival 5100\AA$\,$ continuum data, to construct an optical light curve spanning the same length in time. We also include light curves in the $u$, $B$, $R$, $R1$ and $I$ optical bands, each of length between 3.6 and 6.6 years in time. 

We find both the X-ray and optical emission to be highly variable, but find the amplitude of the optical emission to be smaller than that of the X-rays, even after subtracting the contaminating host galaxy flux falling inside the aperture. The power spectra show that the optical variability power is lower than the X-rays on all time-scales probed. With the current data, the optical PSD is best described by an unbroken power law of slope $\alpha=1.4$, which is much steeper than the low frequency slope of the X-ray PSD ($\alpha_L=1.1$). This suggests that the variations in the different bands do not have a common origin or that they are not related in a simple way. We will address this issue in more detail in a future paper. 

We measured the galaxy flux inside the 15\arcsec$\,$ aperture by decomposition of the optical images, and find the measured fluxes to agree well with a template spectrum of a similar non-active spiral galaxy.

The cross-correlation function between the optical light curve and the X-rays displays two statistically significant peaks. The strongest peak corresponds to the optical variations lagging the X-rays by $1.2^{+1.0}_{-0.3}$~days. This time-scale is consistent with the light travel time to the optical emitting region of the accretion disc, and points to X-ray reprocessing as the source of the fast optical variations. The delays measured between other optical colours and the X-rays are also consistent with this interpretation. However, using a reprocessing model, we find that all the optical variability cannot be accounted for this way. Reprocessing appears to describe the small amplitude, fast variations well, but it fails to reproduce the larger amplitude variations on time-scales of months--years that are also seen in the light curve.

The second significant peak in the CCF is at an optical lag of $38.9^{+2.7}_{-8.4}$~days behind the X-rays. By subtracting a model reprocessed light curve from the observed optical light curve and recalculating the cross-correlation, we can constrain this peak more precisely, viz. $38.9^{+1.2}_{-3.3}$~days. Interpreting this as a light travel time, the corresponding  radius is consistent with the dust sublimation radius in this AGN, which is a theoretical estimate of the inner radius of the dust torus.

As such, we have investigated the origin of the additional optical emission, both in terms of a weak optical contribution from X-ray reprocessing by the dust, and in terms of reflection of optical light from the disc by the dust. We find that neither process can on its own explain the origin of the second peak in the CCF of NGC~4051. It remains possible however, that both these processes may contribute to the optical emission observed. Detailed modeling of the dust emission properties at optical wavelengths, as well as further optical polarimetry will be required to thoroughly test this hypothesis.

In general, the weak (though significant) short time-scale correlation reported here is consistent with a model where the strength of the correlation is determined by the mass of the central black hole and disc accretion rate \citep{uttley03}. When scaled in terms of gravitational radius, a standard accretion disc \citep[e.g.][]{shakura} around a smaller mass black hole, with a given accretion rate, is hotter. The optical emitting region will therefore be further away from the centre and the viscous time scale here will be much longer than near the centre, where the X-rays are emitted. The variations in the different bands may therefore be expected to vary incoherently, weakening the strength of the correlation. Furthermore, the mass-dependence of the disc temperature may also explain why an optical contribution from the torus is detected in this source and not in more massive systems: the cooler discs around more massive systems will emit most of their luminosity at optical wavelengths, dominating over the weak optical signal from the torus. In low mass systems with hotter discs, such as NGC~4051, the peak of the disc emission is shifted to shorter wavelengths, decreasing the fraction of the emission emitted at $V$ band wavelengths and hence increasing the possibility of detecting optical emission from the torus.

\section*{Acknowledgements}
We thank the referee for constructive suggestions which improved this paper. 

We are grateful to the staff of the Rossi X-ray Timing Explorer, the Liverpool Telescope and the Faulkes Telescope for their support of our long term program and also thank Keith Horne and Ed Cackett for their contribution in obtaining this data.

EB gratefully acknowledges financial support in the form of a Stobie-SALT Scholarship from South African National Research Foundation and the University of Southampton. IM$^{\mbox{c}}$H acknowledge support from the STFC under rolling grant PP/D001013/1 and PA acknowledges support from the Natural Science Foundation of China (grants 10773024, 10833002, 10821302, 10825314, 10873030 and 40636031), and National Basic Research Program of China (grant 2009CB824800).
PU acknowledges support from an STFC Advanced Fellowship. 
The work of SGS and NGC is partially supported by the
CosmoMicroPhysics Target Complex Program of the Ukrainian NAS.

\bibliographystyle{mn2e}
\bibliography{library} 

\bsp

\label{lastpage}

\end{document}